\documentclass[preprint,showpacs,preprintnumbers,amssymb,aps,nofootinbib]{revtex4}
\usepackage{amsfonts,color,graphicx,epsfig,graphicx,amsmath,multirow,slashed}
\def\bds{\boldsymbol}

\begin{document}

\title{$X(3860)$ production in association with $J/\psi$ via $e^{+}e^{-}$ annihilation at Belle}
\author{Gui-Yuan Zhang}
\author{Cong Li}
\author{Ying-Zhao Jiang}
\author{Zhan Sun}
\email{zhansun@cqu.edu.cn}
\affiliation{
Department of Physics, Guizhou Minzu University, Guiyang 550025, People's Republic of China.
} 

\date{\today}

\begin{abstract}
In this paper, we study the $X(3860)$ production associated with $J/\psi$ via $e^{+}e^{-}$ annihilation at the next-to-leading-order (NLO) QCD accuracy, within the nonrelativistic QCD (NRQCD) framework. Under the hypothesis of $J^{PC}_{X(3860)}=0^{++}$, the predicted total cross sections agree well with the Belle's measurements at $\sqrt{s}=\Upsilon(4S,5S)$, while the theoretical results given by the $2^{++}$ hypothesis significantly undershoot the data. This is consistent with the Belle's conclusion that the $0^{++}$ hypothesis is favored over the $2^{++}$ hypothesis for $X(3860)$. Despite the agreement of the total cross sections, the NRQCD predictions seem to be incompatible with the Belle-measured $J/\psi$ angular distributions. We also perform the NLO calculations of $\sigma_{e^{+}e^{-} \to J/\psi+X(3940)}$ on the assumption of $J^{PC}_{X(3940)}=0^{-+}$, discovering that the NRQCD predictions that coincide with the light-cone results are in good agreement with the experiment. 
\pacs{12.38.Bx, 12.39.Jh, 14.40.Pq}

\end{abstract}

\maketitle

\section{Introduction}

In 2005, the Belle collaboration firstly observed the charmoniumlike state $X(3915)$ through $B \to J/\psi \omega K$ \cite{Belle:2004lle}, which was subsequently confirmed by the \textit{BABAR} collaboration in the same decay mode \cite{BaBar:2007vxr,BaBar:2010wfc}. In 2012, \textit{BABAR} measured $J^{PC}_{X(3915)}=0^{++}$ in the process of $\gamma \gamma \to X(3915) \to J/\psi \omega$ \cite{BaBar:2012nxg}, following which the $X(3915)$ was identified as the $\chi_{c0}(2P)$ in the 2014 Particle Date Group \cite{ParticleDataGroup:2014cgo}. 

However, this identification has some problems. As the candidate for $\chi_{c0}(2P)$, the $X(3915)$ is expected to dominantly decay into $D\bar{D}$ ($D=$ $D^{0}$ or $D^{+}$), which, however, has not been observed. On the contrary, the decay mode $X(3915) \to J/\psi \omega$ that should largely be suppressed by Okubo-Zweig-Liuka \cite{OZI} is experimentally hunted. In addition, the measured $X(3915)$ width, $20 \pm 5$ Mev, is much smaller than the theoretical expectations \cite{Guo:2012tv}. In 2015, by reanalyzing the \textit{BABAR} data \cite{BaBar:2012nxg}, Zhou \textit{et al.} indicated that the assignment of $J^{PC}_{X(3915)}=2^{++}$ is also permitted, provided the $\lambda=2$ assumption is abandoned \cite{Zhou:2015uva}. In view of these points, the $X(3915)$ is no longer identified as the $\chi_{c0}(2P)$ in the 2016 PDG \cite{ParticleDataGroup:2016lqr}.    

In 2017, by analyzing the process of $e^{+}e^{-} \to J/\psi D\bar{D}$ based on a 980 $\textrm{fb}^{-1}$ data sample, the Belle collaboration observed a new charmoniumlike state $X(3860)$ with a significance of $6.5\sigma$ \cite{Belle:2017egg}. The Belle collaboration concluded that the $J^{PC}=0^{++}$ hypothesis is favored over the $2^{++}$ hypothesis for $X(3860)$. The $X(3860)$ seems to be a better candidate for the $\chi_{c0}(2P)$ charmonium state than the $X(3915)$ \cite{Belle:2017egg,Brambilla:2019esw,Guo:2019twa}. For instance, the mass of $X(3860)$ is measured to be $(3862^{+26+40}_{-32-13})~\textrm{MeV}/c^2$, close to potential model expectations of $m_{\chi_{c0}(2P)}$; the observed decay mode of $X(3860) \to D\bar{D}$ coincides with the theoretical expectations that the charmonium state above the $D\bar{D}$ threshold should primarily decay to $D\bar{D}$; moreover, the measured $X(3860)$ width is $(201^{+154+88}_{-67-82})~\textrm{MeV}$, resembling the theoretical prediction of $\Gamma=(221 \pm 19)$ MeV \cite{Guo:2012tv}. From these perspectives, the $X(3860)$ has been assigned to be $\chi_{c0}(2P)$ by the recent PDG table.

The production of $J/\psi$ in association with positive $C$-parity charmonium via $e^{+}e^{-}$ annihilation is a beneficial process to search for the $C+$ charmonium. The Belle and \textit{BABAR} collaborations have measured the cross sections of $e^{+}e^{-} \to J/\psi+\eta_c(1S,2S),\chi_{c0}(1P)$ \cite{Belle:2002tfa,Belle:2004abn,BaBar:2005nic}, which can reasonably be described by the next-to-leading-order (NLO) predictions \cite{NLO1,NLO2,NLO3,NLO4,NLO5,NLO7} built on the nonrelativistic QCD (NRQCD) framework \cite{Bodwin:1994jh}. In 2005, the Belle collaboration achieved the first detection of $X(3940)$ in $e^{+}e^{-} \to J/\psi+X(3940)$ and measured its cross section \cite{Belle:2005lik}, which is in line with the prediction following the light-cone formalism \cite{Braguta:2006py}. Two years later, the Belle collaboration reported the first observation of $X(4160)$ via $e^{+}e^{-} \to J/\psi+X(4160)$ \cite{Belle:2007woe}; Chao claimed, under the assignment of $X(4160)=\eta_c(4S)~(\textrm{or}~\chi_{c0}(3P))$, that $\sigma_{e^{+}e^{-} \to J/\psi+X(4160)}$ is not small in NRQCD and may be consistent with the experiment \cite{Chao:2007it}.

Inspired by the success of NRQCD in describing the double-charmonium production through $e^{+}e^{-}$ annihilation, in this paper we will use the NRQCD factorization to study the process of $e^{+}e^{-} \to J/\psi+X(3860)$ at the next-to-leading-order (NLO) QCD accuracy. Meanwhile, we will also provide the NLO NRQCD predictions of $\sigma_{e^{+}e^{-} \to J/\psi+X(3940)}$, so as to compare with the existing results given by the light-cone formalism.

The rest of the paper is organized as follows: In Sec. II, we give a description on the calculation formalism. In Sec. III, the phenomenological results and discussions are presented. Section IV is reserved as a summary.

\section{Calculation Formalism}

In our calculations, we choose the hypotheses of $J^{PC}_{X(3860)}=0(2)^{++}$ \cite{Belle:2017egg} and $J^{PC}_{X(3940)}=0^{-+}$ \cite{Braguta:2006py}, which is to say, we treat $X(3860)$ as $\chi_{c0}(2P)$ or $\chi_{c2}(2P)$, and treat $X(3940)$ as $\eta_c(3S)$.

Following the NRQCD formalism, the differential cross section of $e^{+}+e^{-} \to J/\psi+\eta_c(3S),\chi_{cJ}(2P)$ can generally be written as
\begin{eqnarray}
d\sigma=d\hat{\sigma}_{e^{+}e^{-} \to c\bar{c}[n_1]+c\bar{c}[n_2]}\langle \mathcal{O}^{J/\psi}(n_1)\rangle \langle \mathcal{O}^{\eta_c(3S),\chi_{cJ}(2P)}(n_2)\rangle,
\end{eqnarray}
where $d\hat{\sigma}_{e^{+}e^{-} \to c\bar{c}[n_1]+c\bar{c}[n_2]}$ are the perturbative calculable short distance coefficients, denoting the production of a configuration of $c\bar{c}[n_1]$ intermediate state plus $c\bar{c}[n_2]$. According to the above mentioned hypotheses, $n_1=^3S_1$ and $n_2=^1S_0,^3P_{J}$ with $J=0,2$. The universal nonperturbative long distance matrix elements (LDMEs) $\langle \mathcal{O}^{J/\psi}(n_{1})\rangle$ and $\langle \mathcal{O}^{\eta_c(3S),\chi_{cJ}(2P)}(n_{2})\rangle$ stand for the probabilities of $c\bar{c}[n_{1}]$ and $c\bar{c}[n_{2}]$ into $J/\psi$ and $\eta_c(3S)(~\textrm{or}~\chi_{cJ}(2P))$, respectively. 

$d\hat{\sigma}_{e^{+}e^{-} \to c\bar{c}[n_1]+c\bar{c}[n_2]}$ can further be expressed as
\begin{eqnarray}
d\hat{\sigma}_{e^{+}e^{-} \to c\bar{c}[n_1]+c\bar{c}[n_2]}=|\mathcal{M}|^2 d\Pi_{2}=L_{\mu\nu}H^{\mu\nu} d\Pi_{2},
\end{eqnarray}
where $L_{\mu\nu}$ and $H^{\mu\nu}$ are the leptonic and hadronic tensors, respectively, and $d\Pi_{2}$ is the standard two-body phase space.

$L_{\mu\nu}$ reads \cite{NLO7}
\begin{eqnarray}
L_{\mu\nu}=4 \pi \alpha s\left[\mathcal{A}_1(-g_{\mu\nu})+\mathcal{A}_2\frac{{({p}_{J/\psi}})_{\mu}({p}_{J/\psi})_{\nu}}{|{\bds{p}_{J/\psi}}|^2}\right],
\label{lep cur}
\end{eqnarray} 
with
\begin{eqnarray}
\mathcal{A}_1&=&1+\cos^{2}\theta, \nonumber \\
\mathcal{A}_2&=&1-3\cos^{2}\theta,
\end{eqnarray}
where $s=(p_{e^{+}}+p_{e^{-}})^2$, $\bds{p}_{J/\psi}$ is the three momenta of $J/\psi$, and $\theta$ is the angle between $\bds{p}_{J/\psi}$ and the spatial momentum of $e^{-}$ (or $e^{+}$) in the $e^{+}e^{-}$ center-of-mass frame.

Using the leptonic tensor in Eq. (\ref{lep cur}), we write  $d\sigma_{e^{+}e^{-} \to J/\psi+\eta_c(3S),\chi_{cJ}(2P)}$ as
\begin{eqnarray}
\frac{d\sigma}{d\cos\theta}=A+B\cos^2\theta=\kappa\left(\mathcal{C}_1 \mathcal{A}_1+\mathcal{C}_2 \mathcal{A}_2\right),
\label{eq4}
\end{eqnarray}
and then
\begin{eqnarray}
A&=&\kappa(\mathcal{C}_1+\mathcal{C}_2), \nonumber \\
B&=&\kappa(\mathcal{C}_1-3\mathcal{C}_2),  \nonumber \\
\sigma &=& \frac{8}{3}\kappa \mathcal{C}_1.
\label{eqAB}
\end{eqnarray}
The universal factor $\kappa$ follows as
\begin{eqnarray}
\kappa_{J/\psi+\eta_c(3S)} &=& \frac{2 \pi \alpha^2 \alpha_s^2 |R_{1S}(0)|^2 |R_{3S}(0)|^2 \sqrt{s^2-16 m_c^2 s}}{81 m_c^2 s^{2}}, \nonumber \\
\kappa_{J/\psi+\chi_{cJ}(2P)} &=& \frac{2 \pi \alpha^2 \alpha_s^2 |R_{1S}(0)|^2 |R^{'}_{2P}(0)|^2 \sqrt{s^2-16 m_c^2 s}}{27 m_c^2 s^{2}}.
\label{kappa}
\end{eqnarray}
$|R_{1,3S}(0)|$ and $|R^{'}_{2P}(0)|$ are the wave functions at the origin, which can be related to the NRQCD LDMEs by the following formulae:
\begin{eqnarray}
\langle \mathcal O^{J/\psi}(^3S_1) \rangle &=& \frac{9}{2\pi}|R_{1S}(0)|^2, \nonumber  \\
\langle \mathcal O^{\eta_c(3S)}(^1S_0) \rangle &=& \frac{3}{2\pi}|R_{3S}(0)|^2, \nonumber  \\
\langle \mathcal O^{\chi_{cJ}(2P)}(^3P_J) \rangle  &=& (2J+1)\frac{3}{4\pi}|R^{'}_{2P}(0)|^2.
\label{Rs}
\end{eqnarray}

\begin{figure}[!h]
\begin{center}
\hspace{0cm}\includegraphics[width=0.65\textwidth]{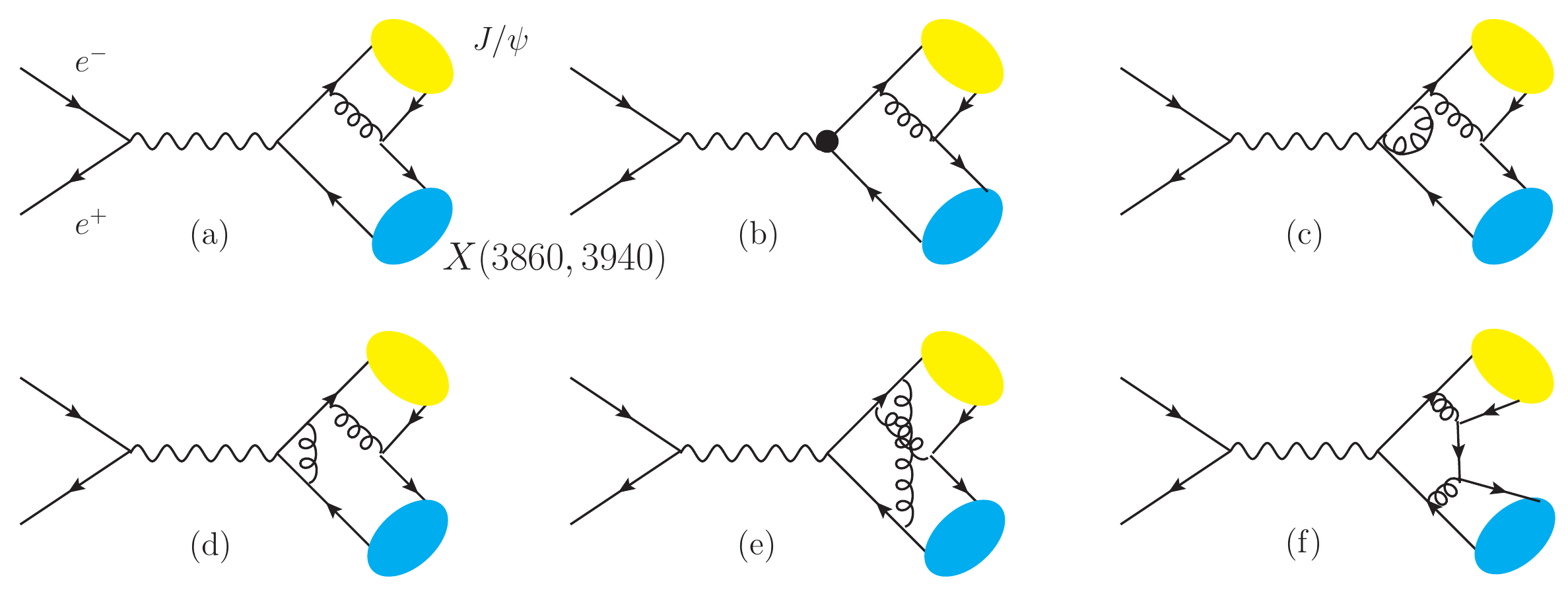}
\caption{\label{fig:Feynman Diagrams}
Representative LO and NLO Feynman diagrams for $e^{+}e^{-} \to J/\psi+X(3860,3940)$.}
\end{center}
\end{figure}

According to Fig. \ref{fig:Feynman Diagrams}(a), the coefficients $\mathcal{C}_{1,2}$ through leading order (LO) in $\alpha_s$ have the following form ($r=4m_c^2/r$)
\begin{itemize}
\item[(i)]
for $J/\psi+\eta_c(3S)$,
\begin{eqnarray}
\mathcal{C}_1&=&-\frac{128 r^3(4r-1)}{m_c^4}, \nonumber \\
\mathcal{C}_2&=&0,
\label{eq5}
\end{eqnarray}
\item[(ii)]
for $J/\psi+\chi_{cJ}(2P)$,
\begin{eqnarray}
\mathcal{C}^{J=0}_1&=&\frac{16 r^2 (144 r^4+152 r^3-428 r^2+182 r+1)}{3 m_c^6}, \nonumber \\
\mathcal{C}^{J=0}_2&=&\frac{16 r^2 (-12 r^2+10 r+1)^2}{3 m_c^6}, \nonumber \\
\mathcal{C}^{J=2}_1&=&\frac{32 r^2 (360 r^4+308 r^3-188 r^2+20 r+1)}{3 m_c^6}, \nonumber \\
\mathcal{C}^{J=2}_2&=&\frac{32 r^2 (360 r^4-96 r^3+4 r^2-4 r+1)}{3 m_c^6}.
\label{eq6}
\end{eqnarray}
\end{itemize}

In the NLO calculations, there are 20 counter-term and 60 one-loop diagrams, as typically shown in Fig. \ref{fig:Feynman Diagrams}(b) and Figs. \ref{fig:Feynman Diagrams}(c)-\ref{fig:Feynman Diagrams}(f), respectively. We utilize the dimensional regularization with $D=4-2\epsilon$ to isolate the ultraviolet (UV) and infrared (IR) divergences. The on-mass-shell (OS) scheme is employed to set the renormalization constants for the $c$-quark mass ($Z_m$) and heavy-quark filed ($Z_2$); the minimal-subtraction ($\overline{MS}$) scheme is adopted for the QCD-gauge coupling ($Z_g$) and the gluon filed $Z_3$. The renormalization constants are taken as \cite{NLO1}

\begin{eqnarray}
\delta Z_{m}^{OS}&=& -3 C_{F} \frac{\alpha_s}{4\pi}N_{\epsilon}\left[\frac{1}{\epsilon_{\textrm{UV}}}+\frac{4}{3}+2\textrm{ln}{2}\right], \nonumber \\
\delta Z_{2}^{OS}&=& - C_{F} \frac{\alpha_s}{4\pi}N_{\epsilon}\left[\frac{1}{\epsilon_{\textrm{UV}}}+\frac{2}{\epsilon_{\textrm{IR}}}+4+6 \textrm{ln}{2}\right], \nonumber \\
\delta Z_{3}^{\overline{MS}}&=& \frac{\alpha_s }{4\pi}(\beta_{0}-2 C_{A})N_{\epsilon}\left[\frac{1}{\epsilon_{\textrm{UV}}}+\textrm{ln}\frac{4m_c^2}{\mu^2}\right], \nonumber \\
\delta Z_{g}^{\overline{MS}}&=& -\frac{\beta_{0}}{2}\frac{\alpha_s }{4\pi}N_{\epsilon}\left[\frac{1} {\epsilon_{\textrm{UV}}}+\textrm{ln}\frac{4m_c^2}{\mu^2}\right], \label{CT}
\end{eqnarray}
where $N_{\epsilon}= \frac{1}{\Gamma[1-\epsilon]}\left(\frac{4\pi\mu^2}{4m_c^2}\right)^{\epsilon}$ is an overall factor, $\gamma_E$ is the Euler's constant, and $\beta_{0}=\frac{11}{3}C_A-\frac{4}{3}T_Fn_f$ is the one-loop coefficient of the $\beta$ function. $n_f(=n_{L}+n_{H})$ represents the number of the active-quark flavors; $n_{L}(=3)$ and $n_{H}(=1)$ denote the number of the light- and heavy-quark flavors, respectively. In ${\rm SU}(3)$, the color factors are given by $T_F=\frac{1}{2}$, $C_F=\frac{4}{3}$, and $C_A=3$.

After including the QCD corrections, we acquire the NLO-level $\mathcal{C}_{1}$ and $\mathcal{C}_{2}$, which can generally be written as
\begin{eqnarray}
\mathcal{C}_{1(2)}^{\textrm{NLO}}=\mathcal{C}_{1(2)}^{\textrm{LO}} \left[ 1+\frac{\alpha_s}{\pi} \left(\frac{1}{2}\beta_{0}\textrm{ln}\frac{\mu_r^2}{4m_c^2}+a_{1(2)}n_{L}+b_{1(2)}n_{H}+c_{1(2)} \right) \right].\label{NLOexp}
\end{eqnarray}
The coefficients $a_{1(2)}$, $b_{1(2)}$, and $c_{1(2)}$, which are functions of $r$ and $m_c$, are summarized in the Appendix (\ref{A1}-\ref{A10}). With the further implementation of Eq. (\ref{eqAB}), one could straightforward obtain the NLO-level $A$, $B$, and $\sigma$.

In our calculations, we use \texttt{FeynArts} \cite{Hahn:2000kx} to generate all the involved Feynman diagrams and the corresponding analytical amplitudes. Then the package \texttt{FeynCalc} \cite{Mertig:1990an} is applied to tackle the traces of the $\gamma$ and color matrices such that the hard scattering amplitudes are transformed into expressions with loop integrals. In the next step, we utilize our self-written $\textit{Mathematica}$ codes with the implementations of \texttt{Apart} \cite{Feng:2012iq} and \texttt{FIRE} \cite{Smirnov:2008iw} to reduce these loop integrals to a set of irreducible Master Integrals, which could be numerically evaluated by using the package \texttt{LoopTools} \cite{Hahn:1998yk}. 

As a cross check, by employing Eqs. (\ref{eqAB}) and (\ref{A1})-(\ref{A3}), one could easily reproduce the $K(=\sigma^{\textrm{NLO}}/\sigma^{\textrm{LO}})$ factors (corresponding to $m_c=1.4$ GeV and $\sqrt{s}=10.58$ GeV) in Refs. \cite{NLO1,NLO2,NLO3,NLO4,NLO5}. We have simultaneously employed the \texttt{FDC} \cite{Wang:2004du} package to independently calculate the QCD corrections, acquiring the same results.
 
\section{Phenomenological results}

In the calculations, $M_{J/\psi} \simeq M_{X(3860)}(\textrm{or}~M_{J/\psi} \simeq M_{X(3940)})=2m_c$ is implicitly adopted to ensure the gauge invariance of the hard scattering amplitude. We choose two typical values for charm-quark mass: 1) $m_c=1.4$ GeV, which corresponds to the one-loop charm quark pole mass \cite{Sang:2020fql}; 2) $m_c$ is identical to the average of $\frac{m_{J/\psi}}{2}$ and $\frac{m_{X(3860,3940)}}{2}$, i.e., $m_c \simeq 1.75$ GeV. The values of wave functions at the origin are correspondingly taken as: for $m_c=1.4$ GeV, $|R_{1S}(0)|^2=0.81\textrm{GeV}^3$, $|R_{3S}(0)|^2=0.455\textrm{GeV}^3$, and $|R^{'}_{2P}(0)|^2=0.102\textrm{GeV}^5$, corresponding to the BT potential model \cite{Buchmuller:1980su,Eichten:1995ch}; as for the large charm quark mass of $m_c=1.75$ GeV, the wave functions are given by the Cornell Potential \cite{Eichten:1995ch}, i.e., $|R_{1S}(0)|^2=1.454\textrm{GeV}^3$, $|R_{3S}(0)|^2=0.791\textrm{GeV}^3$, and $|R^{'}_{2P}(0)|^2=0.186\textrm{GeV}^5$. $\alpha=1/130.9$ \cite{Sang:2020fql} and the two-loop $\alpha_s$ running coupling constant is employed.

\begin{table*}[htb]
\begin{center}
\caption{Comparisons of the predicted total cross sections (in unit: fb) with Belle's measurements of $\sigma_{e^{+}e^{-} \to J/\psi +X(3860)}$. The theoretical uncertainties are caused by the variation of $\mu_r$ in $[2m_c,\sqrt{s}]$ around the central value of $\mu_r=\sqrt{s}/2$. ``$0(2)^{++}$" denotes the hypothesis of $J^{PC}_{X(3860)}=0(2)^{++}$.}
\label{tab: 3860}
\begin{tabular}{|cc|cc|cc|cc|cc|}
\hline
& &\multicolumn{4}{c|}{$m_c=1.4$ GeV} & \multicolumn{4}{c|}{$m_c=1.75$ GeV} \\ \hline
\multicolumn{2}{|c|}{Belle Data} & \multicolumn{2}{c|}{$0^{++}$} & \multicolumn{2}{c|}{$2^{++}$}  & \multicolumn{2}{c|}{$0^{++}$} & \multicolumn{2}{c|}{$2^{++}$} \\ \hline
$\sqrt{s}$ & $\sigma_{\textrm{exp}}$ & $\sigma_{\textrm{LO}}$ & $\sigma_{\textrm{NLO}}$ & $\sigma_{\textrm{LO}}$ & $\sigma_{\textrm{NLO}}$  & $\sigma_{\textrm{LO}}$ & $\sigma_{\textrm{NLO}}$ & $\sigma_{\textrm{LO}}$ & $\sigma_{\textrm{NLO}}$\\ \hline
$\Upsilon(4S)$ & $21.7^{+3.9+2.9}_{-4.3-2.1}$ & $8.30 ^{+4.65}_{-2.48}$ & $15.3 ^{+5.45}_{-3.45}$ & $2.26 ^{+1.26}_{-0.67}$ & $2.63 ^{+0.04}_{-0.31}$  & $13.5 ^{+4.26}_{-4.02}$ & $22.6 ^{+4.18}_{-4.69}$ & $2.34 ^{+0.74}_{-0.70}$ & $2.54 ^{+0.02}_{-0.25}$\\
$\Upsilon(5S)$ & $17.9^{+7.2+2.4}_{-7.3-1.8}$ & $6.74 ^{+3.93}_{-2.00}$ & $12.5 ^{+4.66}_{-2.82}$ & $1.92 ^{+1.12}_{-0.57}$ & $2.25 ^{+0.03}_{-0.26}$  & $11.1 ^{+3.74}_{-3.29}$ & $18.8 ^{+3.70}_{-3.91}$ & $2.06 ^{+0.69}_{-0.61}$ & $2.26 ^{+0.02}_{-0.22}$\\
\hline
\end{tabular}
\end{center}
\end{table*}

We summarize the NRQCD predictions of $\sigma_{e^{+}e^{-} \to J/\psi+X(3860)}$ in Tab. \ref{tab: 3860}.\footnote{Note that, the Belle-measured cross sections at $\sqrt{s}=\Upsilon(1,2,3S)$ and $\sqrt{s}=10.52$ GeV have considerable uncertainties, even exceeding the central values; thus, in our calculations, only the measurements corresponding to $\sqrt{s}=10.58$ GeV ($\Upsilon(4S)$) and $\sqrt{s}=10.87$ GeV ($\Upsilon(5S)$), which have much better precision, are taken into considerations.} Inspecting the data, one can perceive
:
\begin{itemize}
\item[1)]
The QCD corrections provide significant contributions, especially for the $0^{++}$ hypothesis. With the inclusion of the high-order terms, the NLO results show a more steady dependence on the renormalization scale, which can clearly be seen by the uncertainties caused by the variation of $\mu_r$.
\item[2)]
The NLO NRQCD predictions built on the $0^{++}$ hypothesis agree well with the Belle's measurements within uncertainties; however, the theoretical results provided by the $2^{++}$ hypothesis significantly fall short of the data. This is consistent with the Belle's conclusion that the $0^{++}$ hypothesis is favored over the $2^{++}$ hypothesis for $X(3860)$.
\end{itemize} 

\begin{figure}[!h]
\begin{center}
\hspace{0cm}\includegraphics[width=0.32\textwidth]{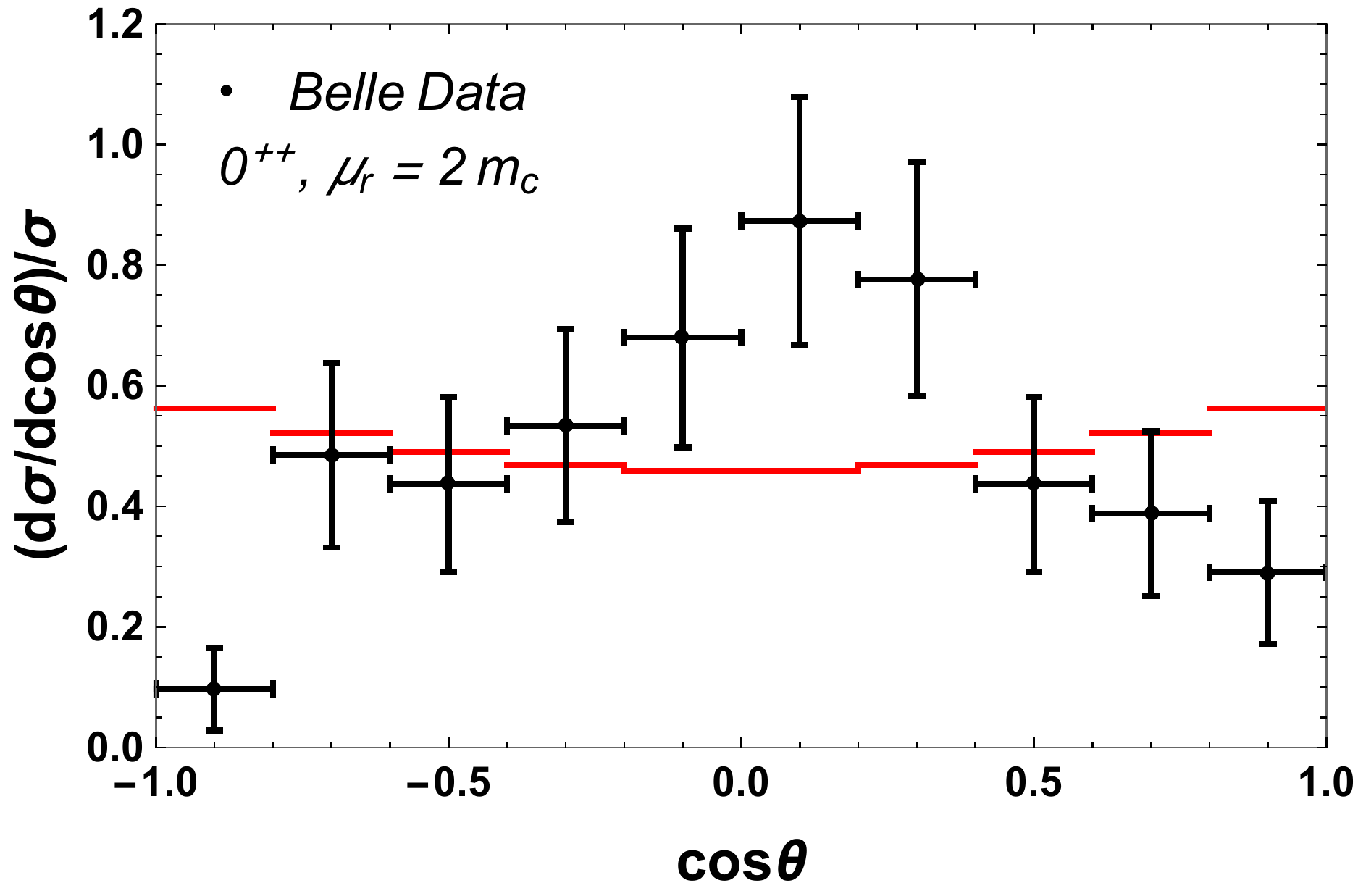}
\hspace{0cm}\includegraphics[width=0.32\textwidth]{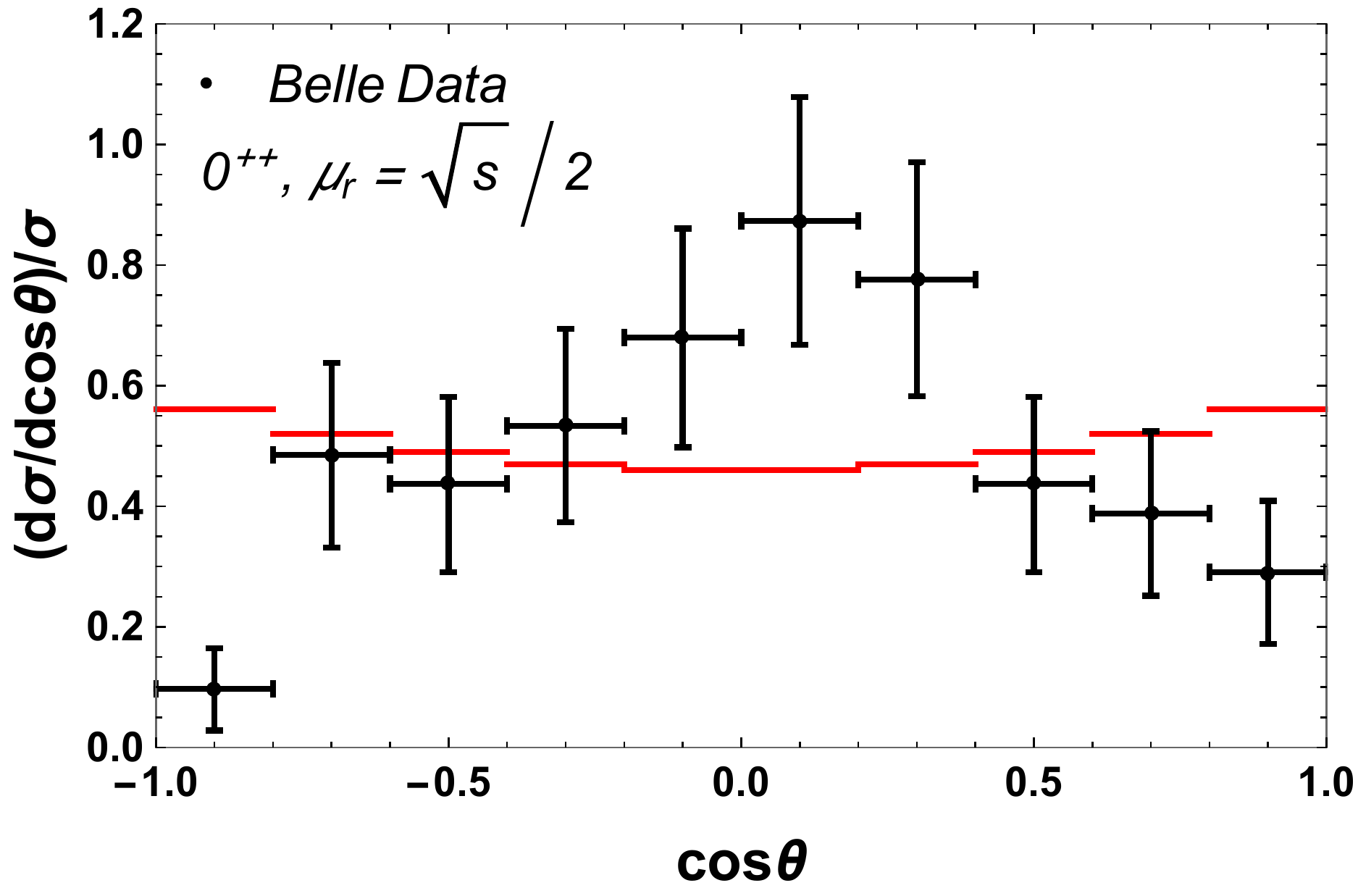}
\hspace{0cm}\includegraphics[width=0.32\textwidth]{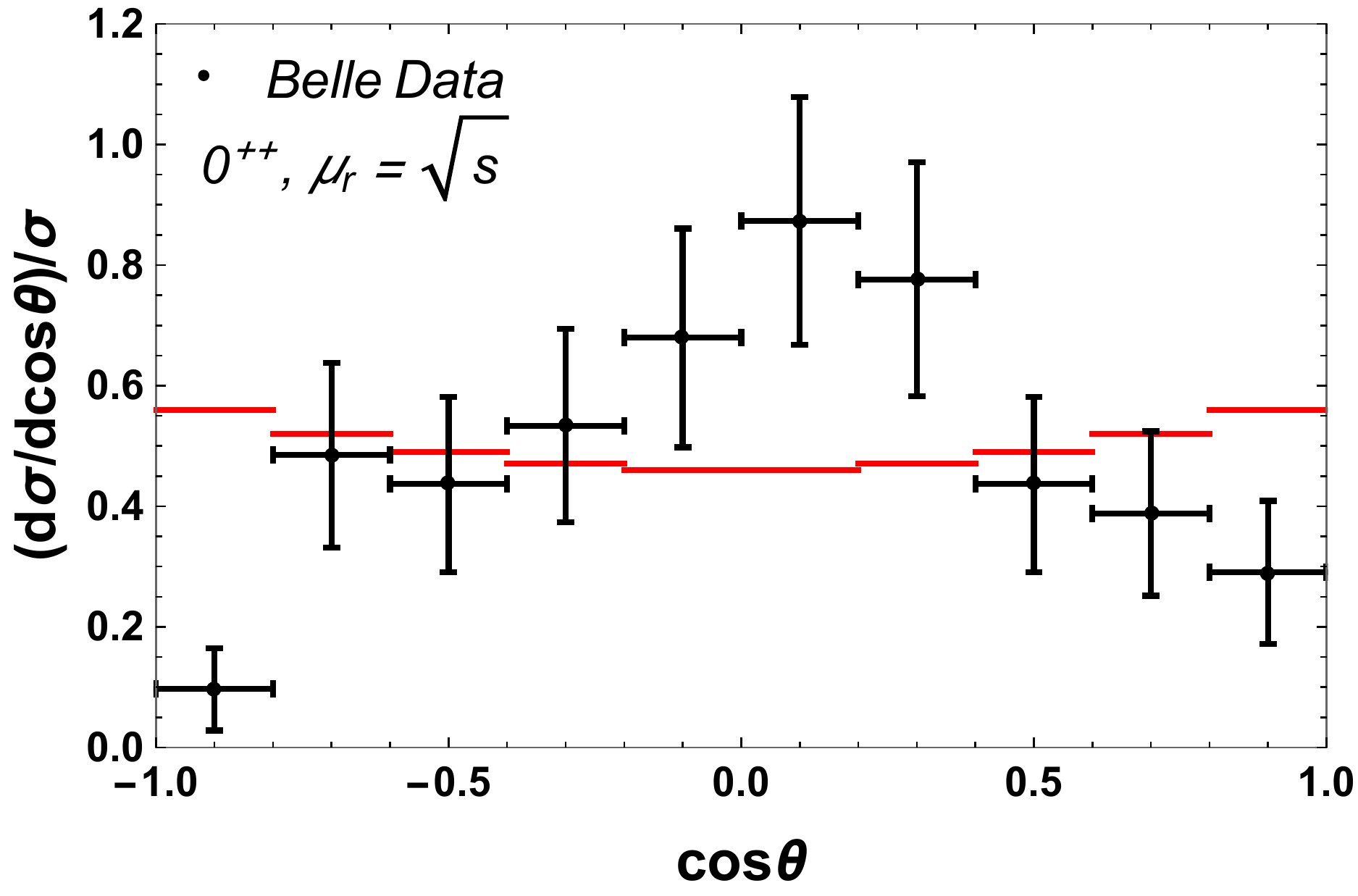}\\
\hspace{0cm}\includegraphics[width=0.32\textwidth]{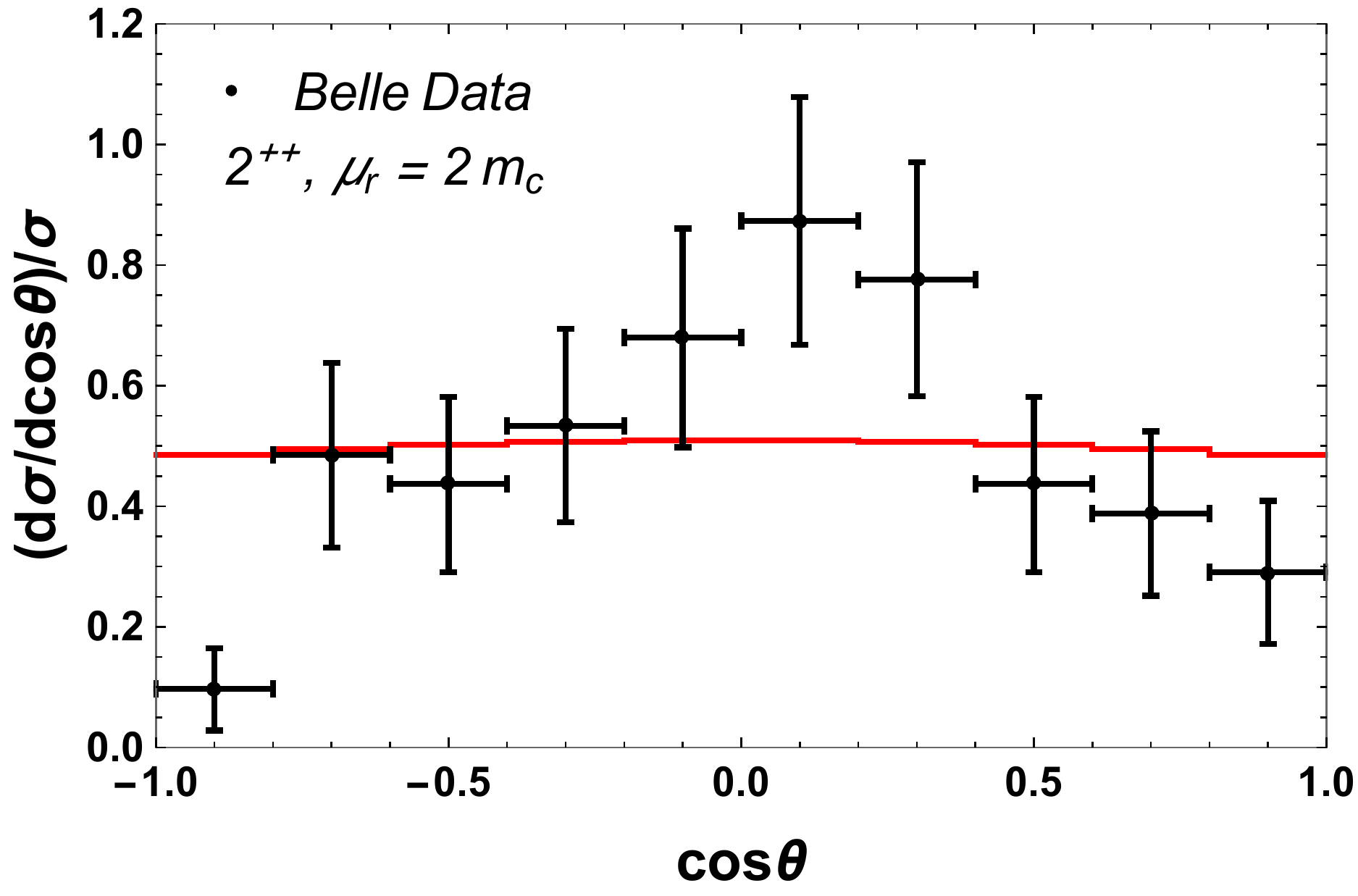}
\hspace{0cm}\includegraphics[width=0.32\textwidth]{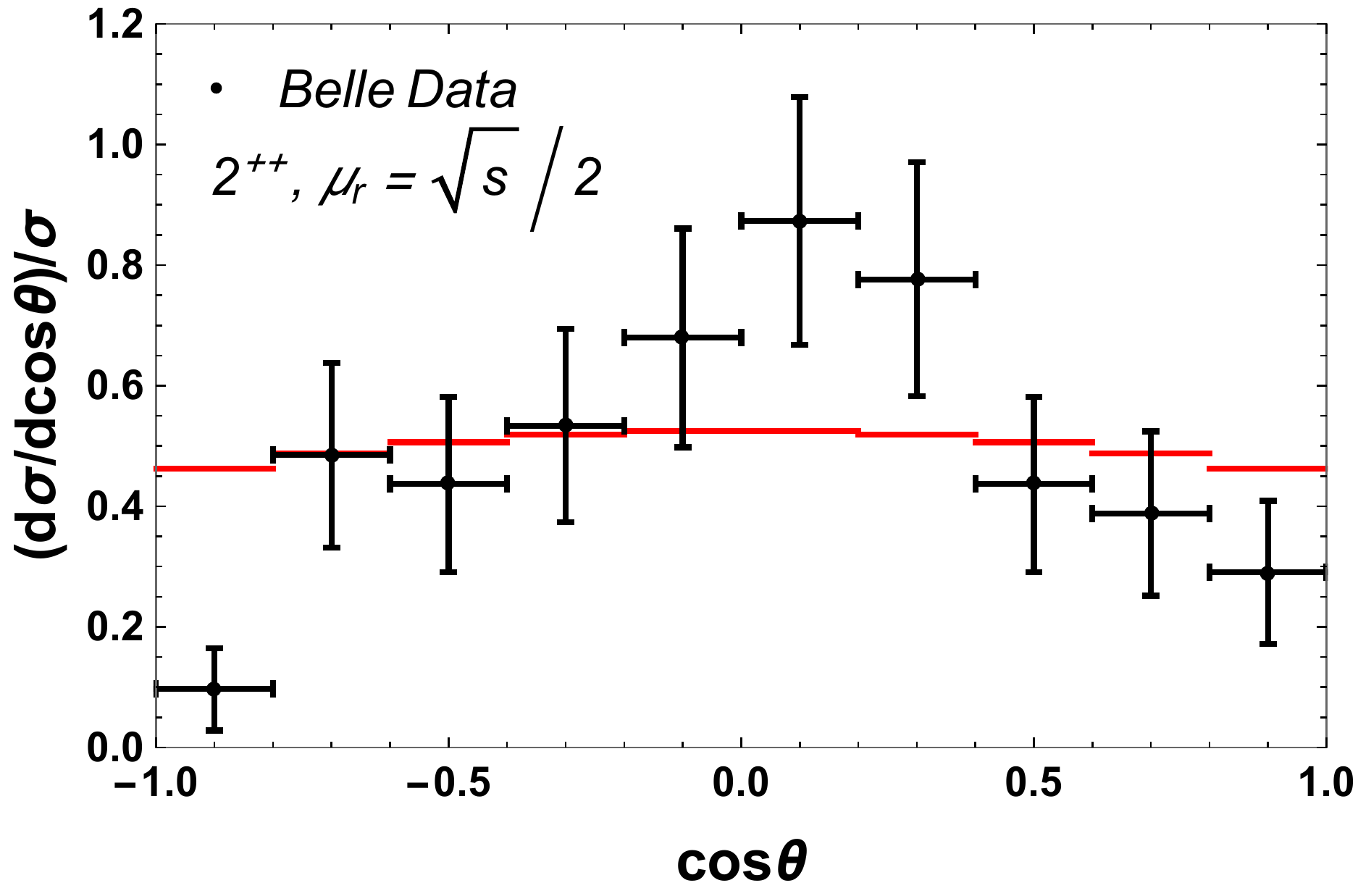}
\hspace{0cm}\includegraphics[width=0.32\textwidth]{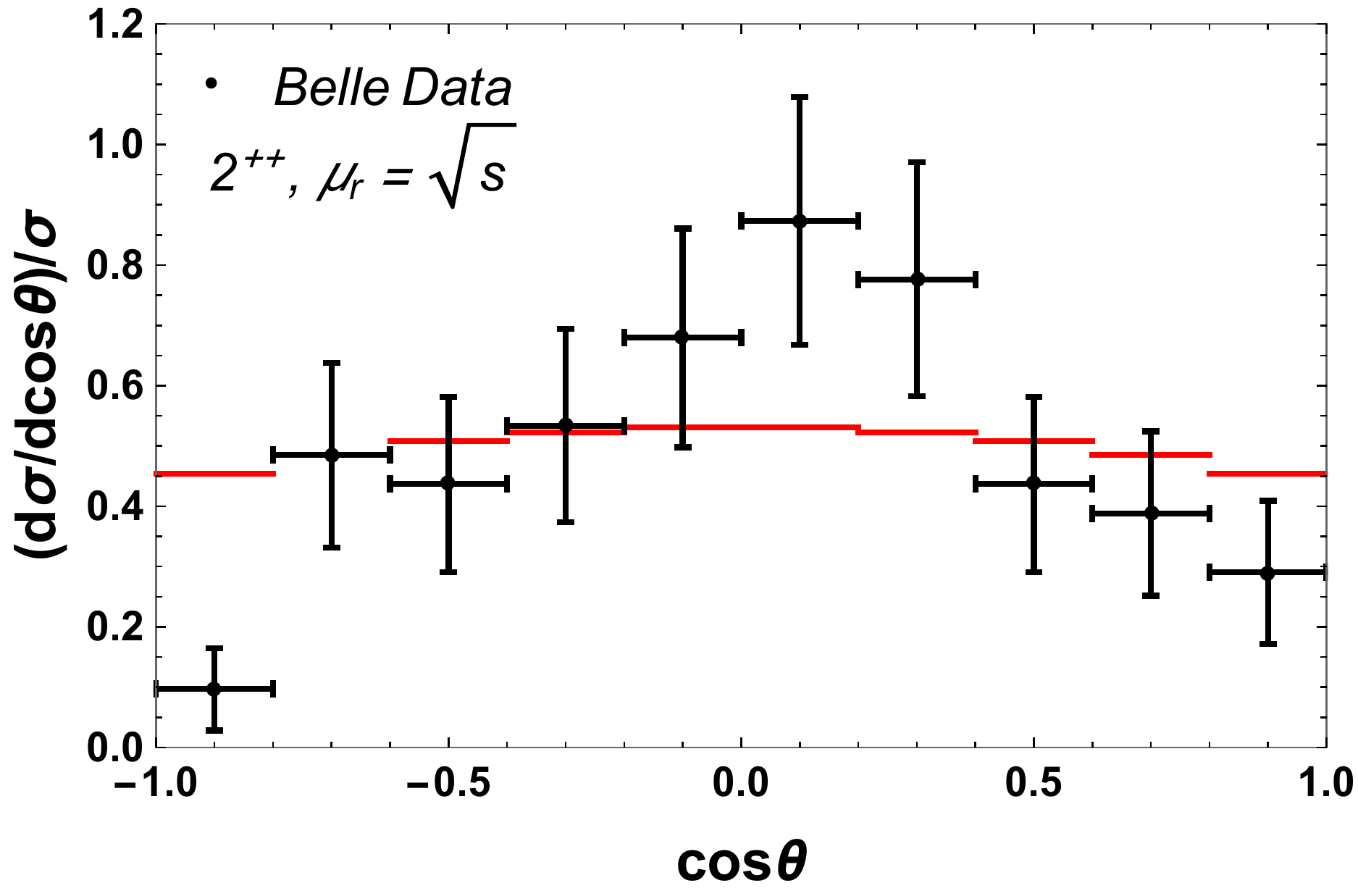}
\caption{\label{fig:dif cross section 1}
Normalized $J/\psi$ differential cross sections as a function of $\cos\theta$ at $\sqrt{s}=10.58$ GeV with $m_c=1.4$ GeV. The solid lines denote the NLO NRQCD results, and ``$0(2)^{++}$" represents the hypothesis of $J^{PC}_{X(3860)}=0(2)^{++}$.}
\end{center}
\end{figure}

\begin{figure}[!h]
\begin{center}
\hspace{0cm}\includegraphics[width=0.32\textwidth]{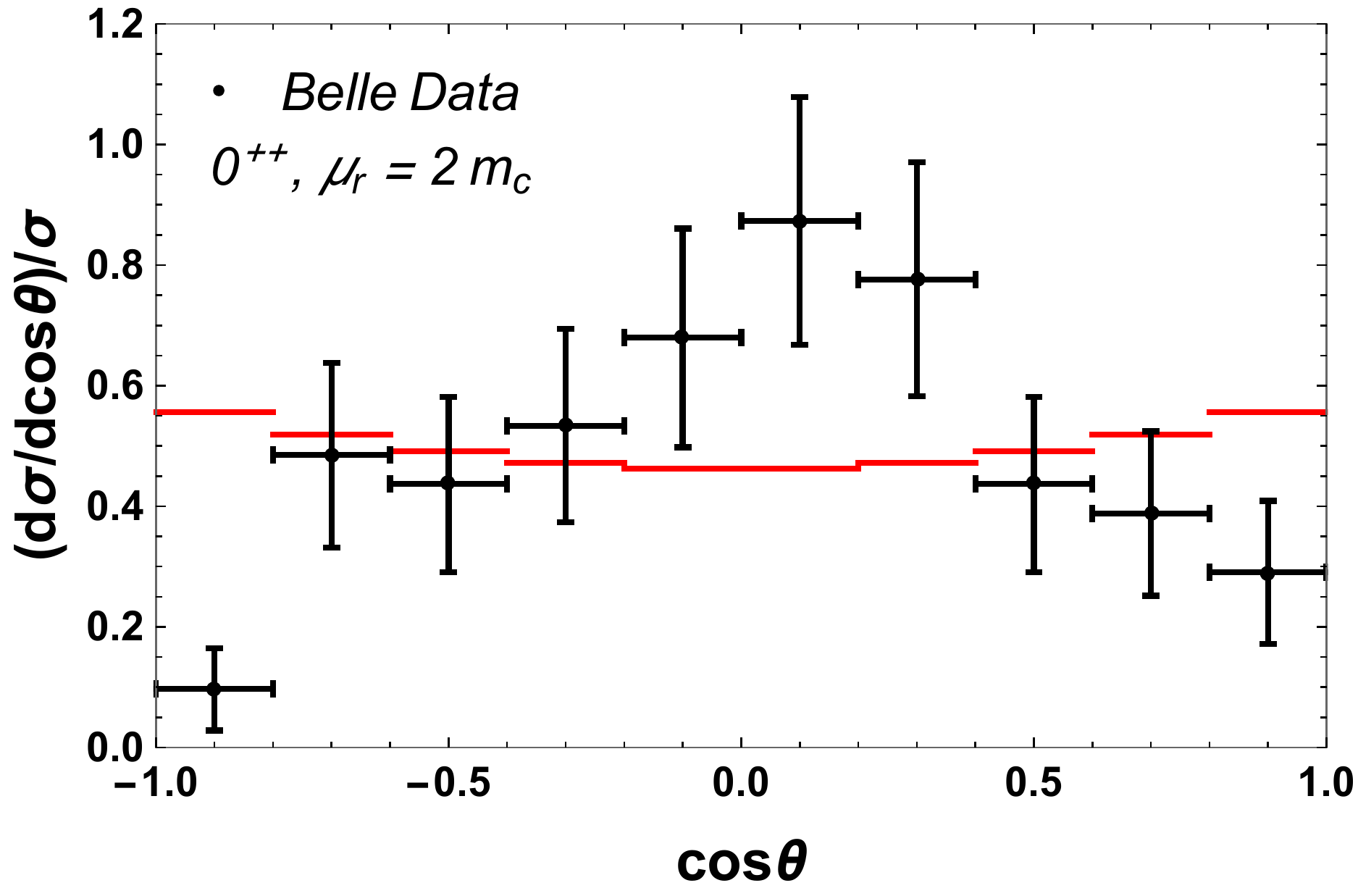}
\hspace{0cm}\includegraphics[width=0.32\textwidth]{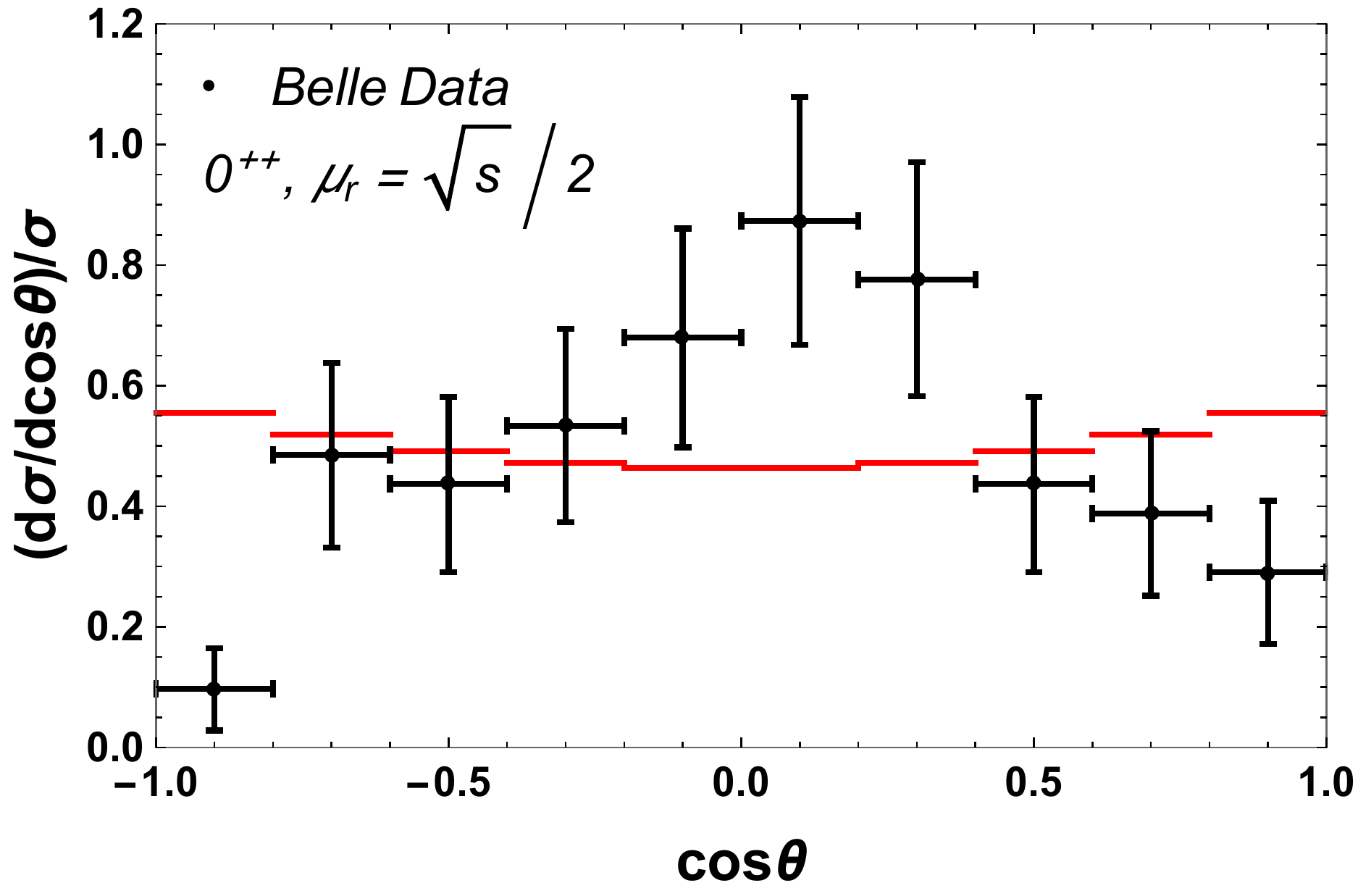}
\hspace{0cm}\includegraphics[width=0.32\textwidth]{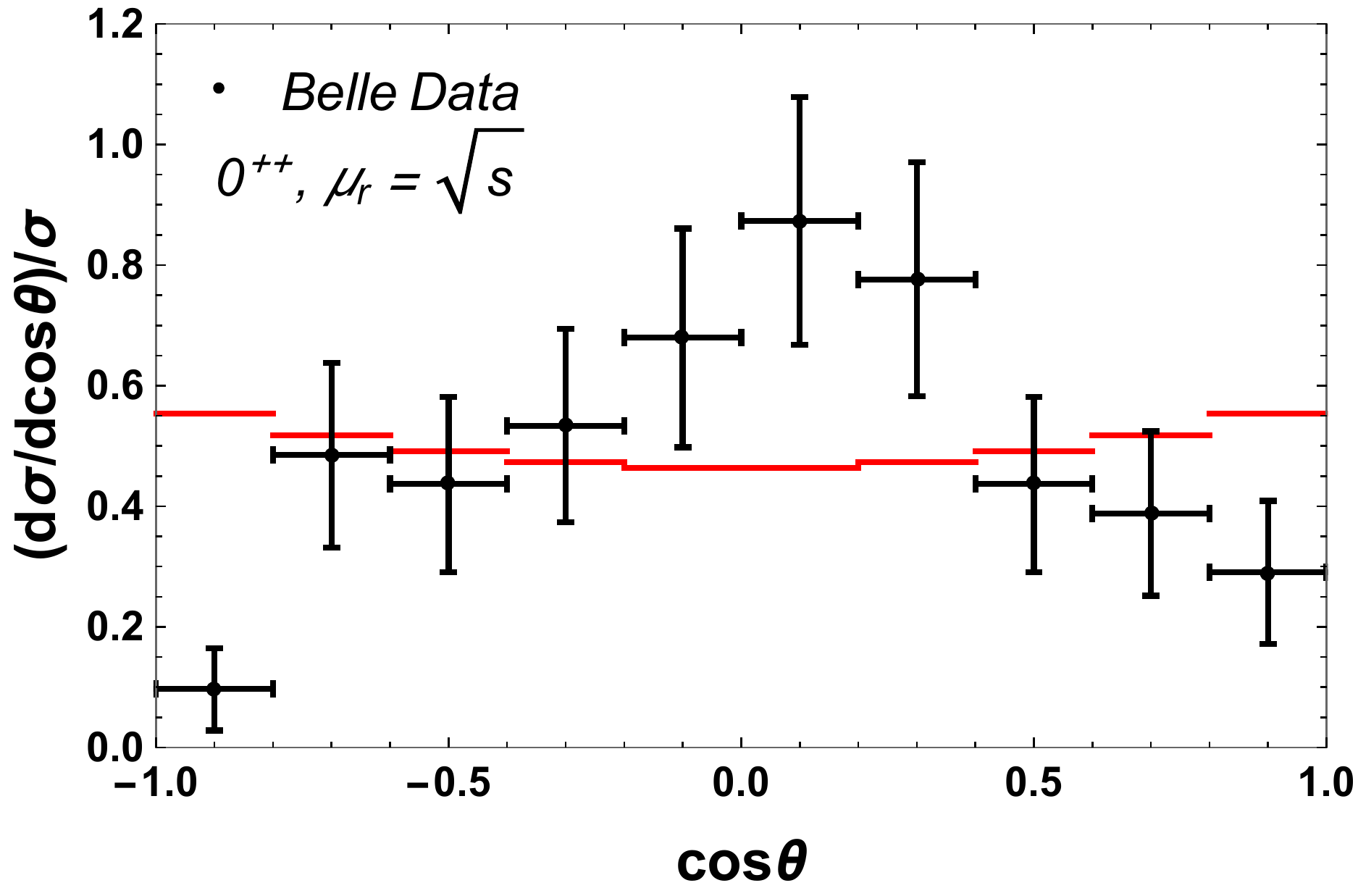}\\
\hspace{0cm}\includegraphics[width=0.32\textwidth]{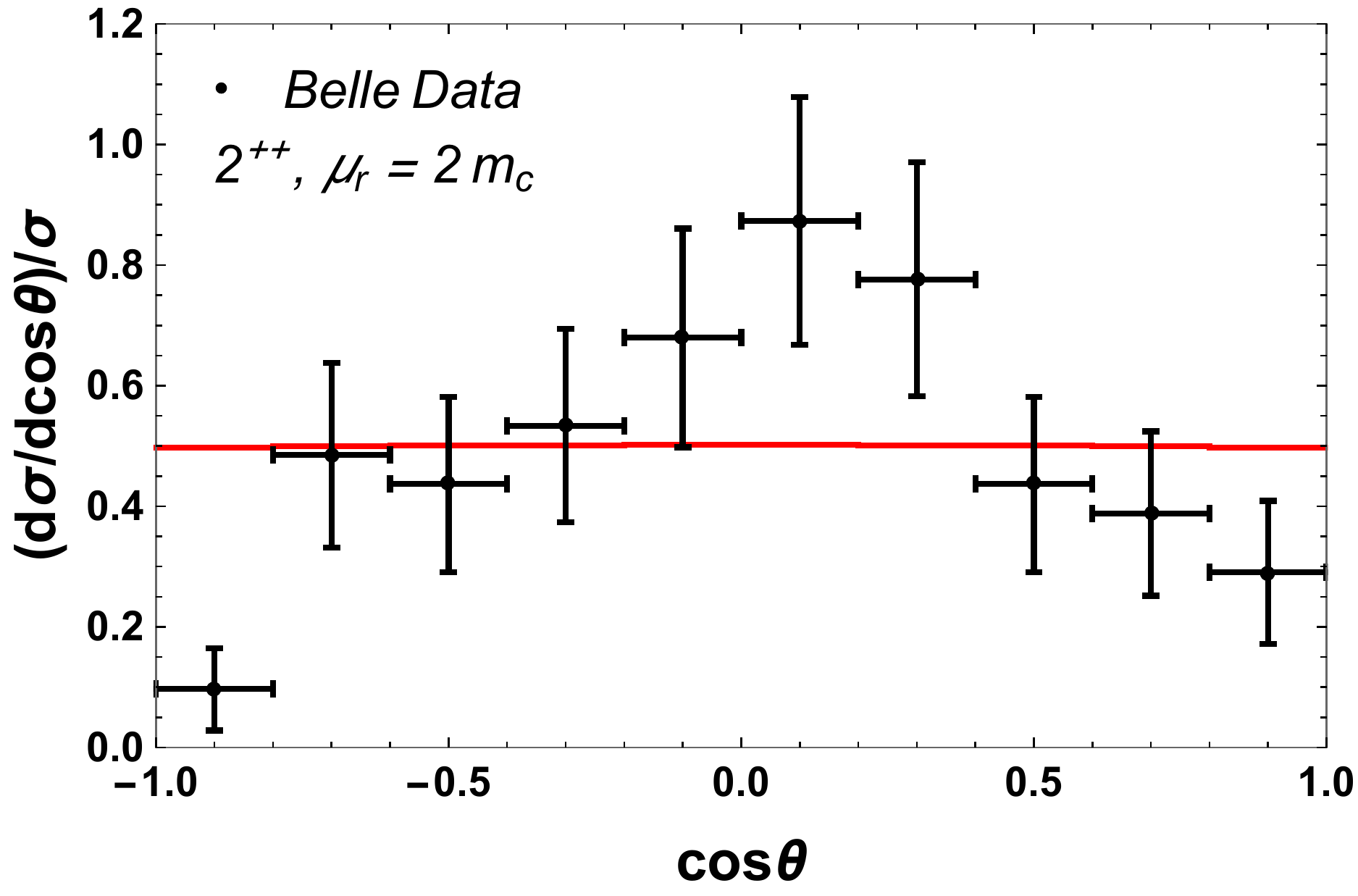}
\hspace{0cm}\includegraphics[width=0.32\textwidth]{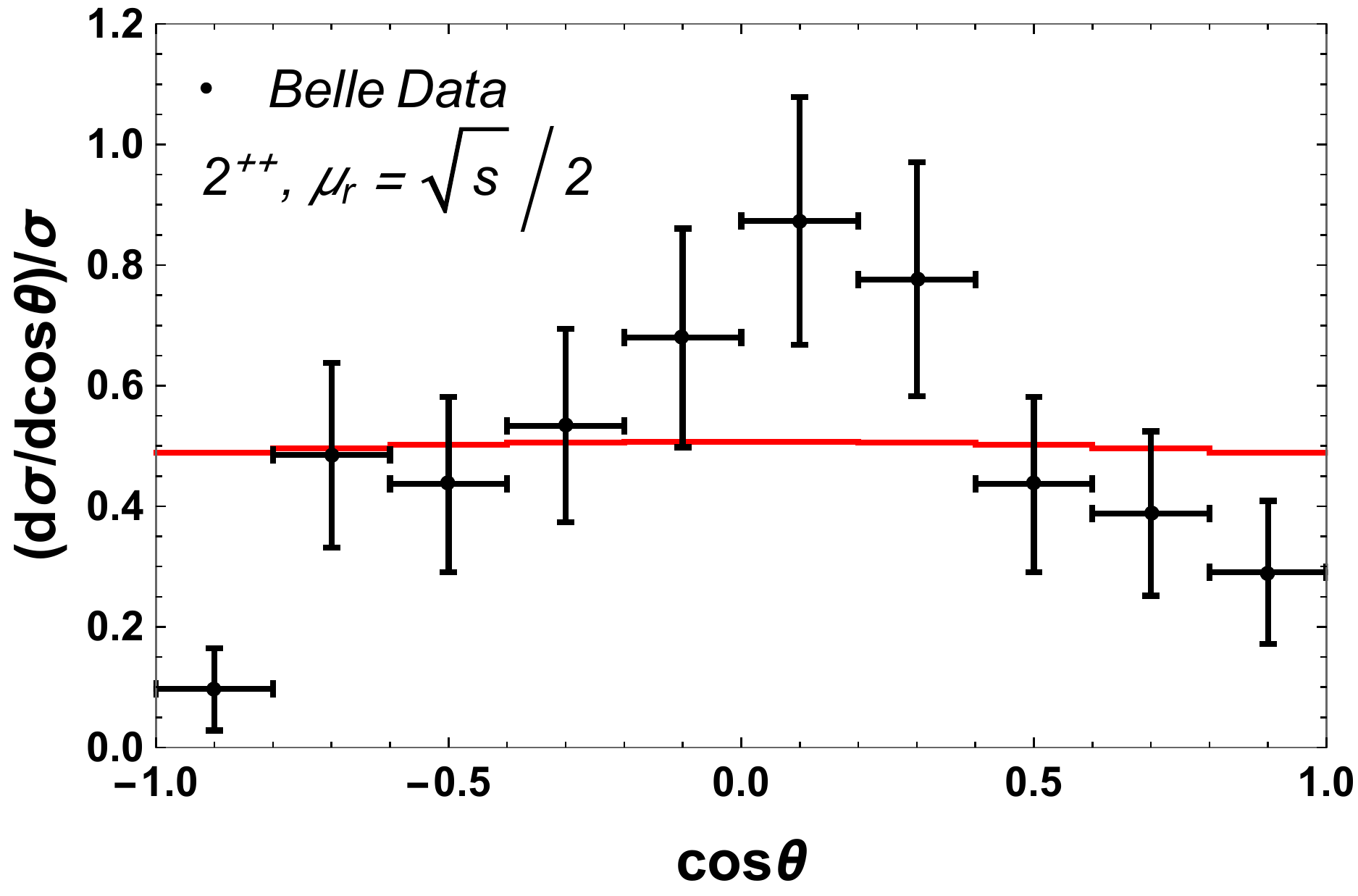}
\hspace{0cm}\includegraphics[width=0.32\textwidth]{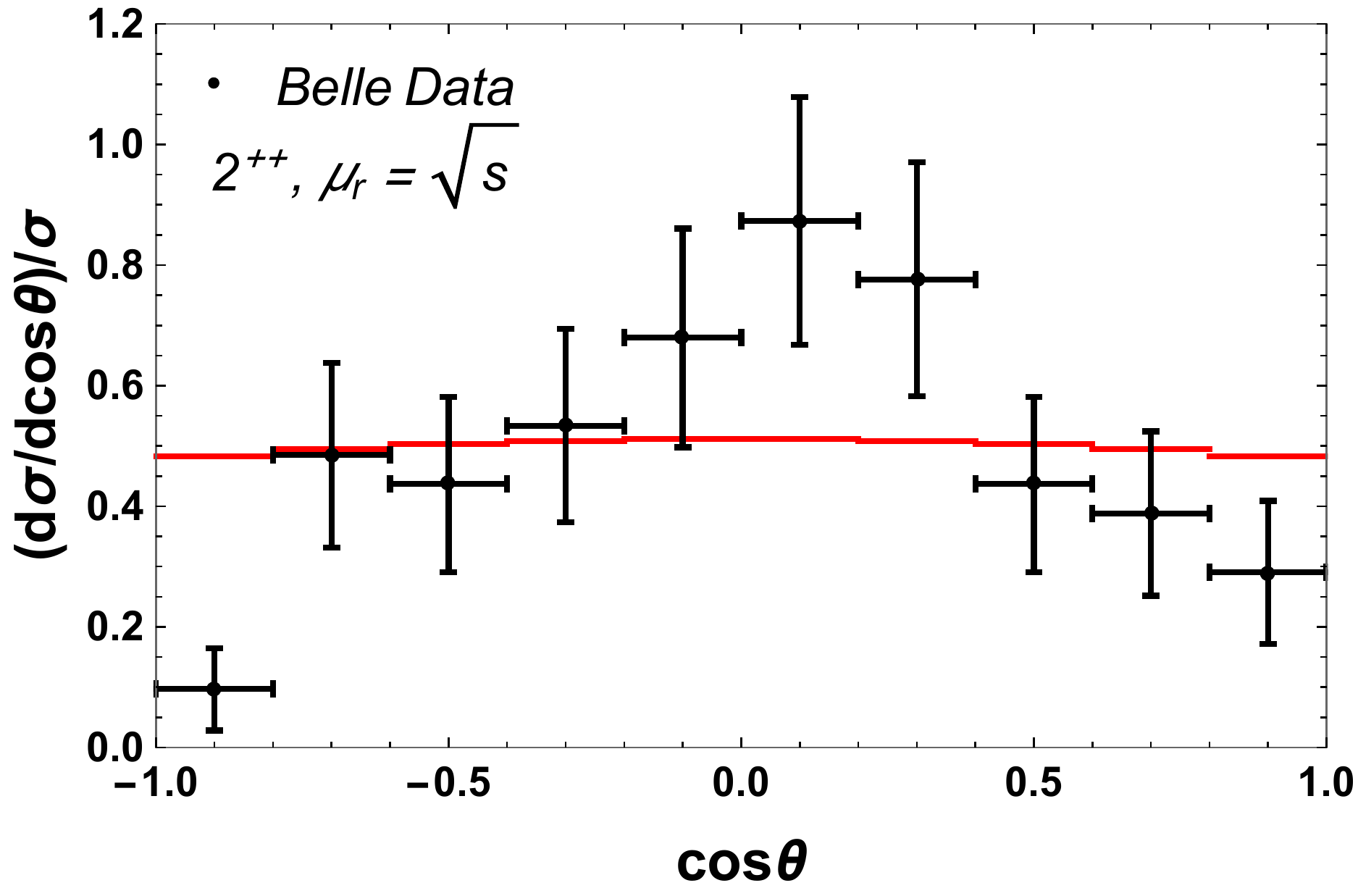}
\caption{\label{fig:dif cross section 2}
Normalized $J/\psi$ differential cross sections as a function of $\cos\theta$ at $\sqrt{s}=10.58$ GeV with $m_c=1.75$ GeV. The solid lines denote the NLO NRQCD results, and ``$0(2)^{++}$" represents the hypothesis of $J^{PC}_{X(3860)}=0(2)^{++}$.}
\end{center}
\end{figure}

In Figs. \ref{fig:dif cross section 1} and \ref{fig:dif cross section 2}, we compare the measurements of the $J/\psi$ angular distributions with the theoretical results of NLO in $\alpha_s$. It appears that the NRQCD predictions could not give reasonable explanations for the experiment. Considering the insensitivity of $\left(\frac{d\sigma}{d\cos\theta}\frac{1}{\sigma}\right)$ on $\mu_r$ (as exemplified by the lines for different $\mu_r$ in Figs. \ref{fig:dif cross section 1} and \ref{fig:dif cross section 2}) and its independence on the choice of the LDMEs, the discrepancies concerning the $J/\psi$ angular distributions seem to hardly be cured, at least at the NLO accuracy. 
\begin{table*}[htb]
\begin{center}
\caption{Comparisons of the predicted total cross sections (in unit: fb) with Belle's measurements of $\sigma_{e^{+}e^{-} \to J/\psi +X(3940)}$. The theoretical uncertainties are caused by the variation of $\mu_r$ in $[2m_c,\sqrt{s}]$ around the central value of $\mu_r=\sqrt{s}/2$.}
\label{tab: 3940}
\begin{tabular}{|cc|cc|cc|cc|cc|}
\hline
\multicolumn{2}{|c|}{Belle Data} &\multicolumn{2}{c|}{$m_c=1.4$ GeV} & \multicolumn{2}{c|}{$m_c=1.75$ GeV} \\ \hline
$\sqrt{s}$ & $\sigma_{\textrm{exp}}$ & $\sigma_{\textrm{LO}}$ & $\sigma_{\textrm{NLO}}$ & $\sigma_{\textrm{LO}}$ & $\sigma_{\textrm{NLO}}$\\ \hline
$\Upsilon(4S)$ & $10.6 \pm 2.5 \pm 2.4$ & $2.50 ^{+1.40}_{-0.75}$ & $5.49 ^{+2.49}_{-1.41}$ & $5.39 ^{+1.70}_{-1.61}$ & $10.5 ^{+2.43}_{-2.49}$\\
\hline
\end{tabular}
\end{center}
\end{table*}

At last, by identifying $X(3940)$ as $\eta_c(3S)$ \cite{Braguta:2006py}, we provide the comparisons of the NRQCD predictions of $\sigma_{e^{+}e^{-} \to J/\psi+X(3940)}$ with the Belle's data in Tab. \ref{tab: 3940}. One can observe that the NLO NRQCD results, which exhibit a good consistence with the light-cone results (i.e., $11 \pm 3$ fb) \cite{Braguta:2006py}, are in good agreement with the measurements within the respective theory errors.
\section{Summary}

In this manuscript, we applied the NRQCD factorization to study the production of $X(3860)$ plus $J/\psi$ via $e^{+}e^{-}$ annihilation at the NLO QCD accuracy. We found, by assuming $J^{PC}_{X(3860)}=0^{++}$, that the NRQCD predictions of $\sigma_{e^{+}e^{-} \to J/\psi+X(3860)}$ match the Belle's measurements at $\sqrt{s}=\Upsilon(4S,5S)$ adequately; however, the results given by the $2^{++}$ hypothesis largely undershoot the data, which coincide with the Belle's conclusion that the $0^{++}$ hypothesis is favored over the $2^{++}$ hypothesis for $X(3860)$. In spite of the agreement of the total cross sections, the NRQCD predictions seem hardly to describe the Belle-measured $J/\psi$ angular distributions. With the interpretation of $X(3940)$ as $\eta_c(3S)$, we simultaneously carried out the NLO calculations of $\sigma_{e^{+}e^{-} \to J/\psi+X(3940)}$, finding the NRQCD predictions to be consistent with the light-cone results and be in good agreement with the experiment.

\section{Acknowledgments}
\noindent{\bf Acknowledgments}:
This work is supported in part by the Natural Science Foundation of China under the Grant No. 12065006, and by the Project of GuiZhou Provincial Department of Science and Technology under Grant No. QKHJC[2019]1167. and No. QKHJC[2020]1Y035.\\

\appendix

\section{}
In this section, we list the numerical values of the coefficients $a_{1(2)}$, $b_{1(2)}$, and $c_{1(2)}$ in Eq. (\ref{NLOexp}).

For $\sqrt{s}=10.58$ GeV:
\begin{itemize}
\item[I.]$m_c=1.4$ GeV,
\begin{itemize}
\item[1)]$J/\psi+\eta_c(3S)$,
\begin{eqnarray}
&&a_1=-0.1314,~b_1=-0.3046,~c_1=13.057, \nonumber \\
&&a_2=b_2=c_2=0,\label{A1}
\end{eqnarray}
\item[2)]$J/\psi+\chi_{c0}(2P)$,
\begin{eqnarray}
&&a_1=-0.2159,~b_1=-0.4392,~c_1=8.1957, \nonumber \\
&&a_2=-0.2981,~b_2=-0.5702,~c_2=7.5214,\label{A2}
\end{eqnarray}
\item[3)]$J/\psi+\chi_{c2}(2P)$,
\begin{eqnarray}
&&a_1=-0.4112,~b_1=-0.7505,~c_1=-0.8632, \nonumber \\
&&a_2=-0.4534,~b_2=-0.8179,~c_2=-2.4466.\label{A3}
\end{eqnarray}
\end{itemize}
\item[II.]$m_c=1.75$ GeV,
\begin{itemize}
\item[1)]$J/\psi+\eta_c(3S)$,
\begin{eqnarray}
&&a_1=-0.2802,~b_1=-0.5710,~c_1=12.060, \nonumber \\
&&a_2=b_2=c_2=0,
\end{eqnarray}
\item[2)]$J/\psi+\chi_{c0}(2P)$,
\begin{eqnarray}
&&a_1=-0.3463,~b_1=-0.7051,~c_1=8.3116, \nonumber \\
&&a_2=-0.3973,~b_2=-0.8086,~c_2=7.9286,
\end{eqnarray}
\item[3)]$J/\psi+\chi_{c2}(2P)$,
\begin{eqnarray}
&&a_1=-0.5436,~b_1=-1.1055,~c_1=0.5897, \nonumber \\
&&a_2=-0.5856,~b_2=-1.1908,~c_2=-0.4977.
\end{eqnarray}
\end{itemize}
\end{itemize}

For $\sqrt{s}=10.87$ GeV:
\begin{itemize}
\item[I.]$m_c=1.4$ GeV,
\begin{itemize}
\item[1)]$J/\psi+\chi_{c0}(2P)$,
\begin{eqnarray}
&&a_1=-0.1998,~b_1=-0.4107,~c_1=8.1818, \nonumber \\
&&a_2=-0.2856,~b_2=-0.5442,~c_2=7.4574,
\end{eqnarray}
\item[2)]$J/\psi+\chi_{c2}(2P)$,
\begin{eqnarray}
&&a_1=-0.3946,~b_1=-0.7139,~c_1=-1.0438, \nonumber \\
&&a_2=-0.4365,~b_2=-0.7791,~c_2=-2.6764.
\end{eqnarray}
\end{itemize}
\item[II.]$m_c=1.75$ GeV,
\begin{itemize}
\item[1)]$J/\psi+\chi_{c0}(2P)$,
\begin{eqnarray}
&&a_1=-0.3308,~b_1=-0.6696,~c_1=8.2981, \nonumber \\
&&a_2=-0.3856,~b_2=-0.7773,~c_2=7.8893,
\end{eqnarray}
\item[2)]$J/\psi+\chi_{c2}(2P)$,
\begin{eqnarray}
&&a_1=-0.5283,~b_1=-1.0571,~c_1=0.4154, \nonumber \\
&&a_2=-0.5707,~b_2=-1.1403,~c_2=-0.7404.\label{A10}
\end{eqnarray}
\end{itemize}
\end{itemize}

\end{document}